\documentstyle[aps,twocolumn]{revtex}
\begin{document}
\draft
\title{Long--time relaxation of current in a 2D weakly disordered conductor.}
\author{ Alexander D. Mirlin}
\address{Institut f\"{u}r Theorie der Kondensierten Materie,
  Universit\"{a}t Karlsruhe, 76128 Karlsruhe, Germany}
\address{
and  Petersburg Nuclear Physics Institute, 188350 Gatchina, St.Petersburg,
Russia.}
\date{\today}
\maketitle
\tighten
\begin{abstract}
The long-time relaxation of the average conductance in a 2D mesoscopic
sample is studied within the method recently suggested by Muzykantskii and
Khmelnitskii and based on a saddle-point approximation to
the supermatrix $\sigma$--model. The obtained far asymptotics is
in perfect agreement with the result of renormalization group
treatment by Altshuler, Kravtsov and Lerner.
\end{abstract}
\narrowtext

In the recent paper \cite{MK}, Muzykantskii and Khmelnitskii (MK)
considered the relaxation phenomena in disordered conductors in the
framework of the supersymmetric $\sigma$--model approach. They
suggested a nice idea that the long-time asymptotics of the
conductance $G(t)$ is governed by a
non-trivial saddle point of the $\sigma$--model. Their original goal
was to reproduce in a more direct way the result of Altshuler,
Kravtsov and Lerner (AKL) \cite{AKL}, who found the logarithmically
normal (LN) ``tail'' in the time dispersion in two and $(2+\epsilon)$
dimensions. However, MK found in 2D a different, power-law decay for
moderately large times. They put forward a hypothesis
 that the LN asymptotics could hold for longer times. Here I will show
that this is indeed the case,
and that this result can be obtained via the method developed by MK.

Following MK, I consider a 2D disk-shaped sample of a radius $R$. I
will consider the unitary symmetry (broken time reversal invariance)
in course of the calculations. For systems of the orthogonal and
symplectic symmetries, the treatment is completely analogous, and I
simply present the corresponding results in the end of the paper.
The problem can be described by the $\sigma$--model with the action
\cite{efe}
\begin{equation}
S=-{\pi\nu\over 4}\int d^2 r\,\mbox{Str}[D(\nabla Q)^2+2i\omega\Lambda Q]
\label{1a}
\end{equation}
Here $Q(\bbox{r})$ is $4\times 4$ supermatrix field, $D$ is the
diffusion constant, $\nu$ the density of states, $\omega$ the
frequency, $\mbox{Str}$ denotes the supertrace, and
$\Lambda=\mbox{diag}(1,1,-1,-1)$. The saddle point equation of MK
reads:
\begin{equation}
\Delta_L\theta+\gamma^2\sinh\theta=0\ ,
\label{2a}
\end{equation}
where $\theta(\bbox{r})$ is the ``non-compact angle'' parametrizing
the $\sigma$-model field $Q(\bbox{r})$, $\Delta_L$ is the Laplace
operator and $\gamma^2=i\omega/D$. It should be supplemented by the
boundary conditions on the boundary with leads
\begin{equation}
\theta|_{\mbox{leads}}=0
\label{3a}
\end{equation}
and on insulating boundary
\begin{equation}
\nabla_{\bbox{n}} \theta|_{\mbox{insulator}}=0\ ,
\label{4a}
\end{equation}
where $\nabla_{\bbox{n}}$ denotes the normal derivative.

We can consider the two leads attached to the disk boundary
to be of almost semicircular shape, with relatively narrow insulating
intervals between them. Then we can approximate the boundary
conditions by using eq.(\ref{3a}) for all the boundary, as it was done
by MK. In fact, in view of the logarithmic dependence of the saddle
point action on $R$ (see below), the result should not depend to the
leading aproximation on the specific shape of the sample and  the
leads attached. With the rotationally invariant form of the boundary
condition, the minimal action corresponds to the function $\theta$
depending on the radius $r$ only. We get therefore the radial equation
\begin{equation}
\theta'' + \theta'/r+\gamma^2\sinh\theta=0\ ;\qquad 0\le r\le R
\label{1}
\end{equation}
(the prime denotes the derivative $d/dr$)
with the boundary conditions:
\begin{eqnarray}
&&\theta(R)=0  \label{6a}\ ,\\
&& \theta'(0)=0 \label{7a}
\end{eqnarray}
The condition (\ref{7a}) follows from the requirment of analyticity
of the field in the disk center.

Assuming that characteristic values of $\theta$ satisfy the condition
$\theta\gg 1$
(we will find below the corresponding restriction on the time $t$),
one can replace
$\sinh\theta$ by $e^\theta/2$. Eq.(\ref{1}) can be then easily
integrated, and its general solution reads:
\begin{equation}
e^{\theta(r)}={4C_1^2\over\gamma^2}
{C_2r^{C_1-2}\over (C_2 r^{C_1}+1)^2}\ ,
\label{g}
\end{equation}
with two integration constants $C_1$ and $C_2$.
 To satisfy the boundary condition (\ref{7a}), we have to choose
$C_1=2$. Furthermore, the above assumption $\theta(0)\gg1$ implies
that $4C_2/\gamma^2\gg1$. Therefore, the second boundary condition
(\ref{6a}) is satisfied if $C_2\simeq (4/\gamma R^2)^2$, and the
solution can be written in the form
\begin{equation}
e^{\theta(r)}\simeq[(r/R)^2 + (\gamma R/4)^2]^{-2}
\label{3}
\end{equation}
Using now the self-consistency equation of MK,
\begin{equation}
2\pi\int_0^R dr\,r(\cosh\theta-1)=t/\pi\nu\ ,
\label{4}
\end{equation}
one finds $\gamma^2=8\pi^2\nu/t$. Finally, the action
\begin{equation}
S\simeq\pi^2\nu D\int dr\,r(\theta^{'2}-\gamma^2 e^\theta)
\label{5}
\end{equation}
is equal on the saddle point (\ref{3}) to
\begin{equation}
S\simeq 8\pi^2\nu D\ln(t\Delta)\ ,
\label{5b}
\end{equation}
where $\Delta=1/(\nu\pi R^2)$ is the mean level spacing.
Eq.(\ref{5}) coincides exactly with the result of MK. This
consideration is valid provided $\theta'(r)<l^{-1}$ on the saddle
point solution, which is the condition of the applicability of the
diffusion approximation (here $l$ is the mean free path). In
combination with the assumption $\theta(0)\gg 1$ this means that $1\ll
t\Delta\ll (R/l)^2$.

Now I consider the ultra-long-time region, $t\gg \Delta^{-1}
(R/l)^2$. In order to support the applicability of the diffusion
approximation, we should search for a function $\theta(r)$
minimizing the action with an additional restriction $\theta'\le l^{-1}$.
Since the derivative has a tendency to increase in the vicinity of $r=0$,
the restriction can be implemented via replacing the boundary conditions
(\ref{7a}) by $\theta'(r_*)=0$, where the parameter
$r_*$ will be specified below. The solution reads now:
\begin{equation}
e^{\theta(r)}=\frac { (r/R)^{C-2}} {[(r/R)^C + {C+2\over C-2}
(r_*/R)^C]^2}\ ;\ \ r_*\le r\le R
\label{6}
\end{equation}
The function $\theta(r)$ is ment as being constant within the vicinity
$|r|\le r_*$ of the disk center.
The condition $\theta'\le l^{-1}$ yields $r_*\sim lC$.
It is important to note that the result does not depend on details
of the cut-off procedure. For example one gets the same results if
one chooses the boundary condition in the form $\theta'(r_*)=1/l$.
The crucial point  is that the maximal derivative $\theta'$ should not
exceed $1/l$. The constant $C$ is to be found from the
self-consistency  equation
(\ref{4}) which can be reduced to the following form:
\begin{equation}
\left({R\over r_*}\right)^C= {2t\over\pi^2\nu R^2} {C^2\over C-2}
\label{8a}
\end{equation}
Neglecting corrections of the $\ln(\ln\cdot)$ form, we find
\begin{equation}
C\simeq {\ln (t\Delta) \over \ln (R/r_*)}\simeq
{\ln (t\Delta) \over \ln (R/l)}
\label{7}
\end{equation}
The action (\ref{5}) is then equal to
\begin{equation}
S\simeq\pi^2\nu D (C+2)^2\ln (R/r_*)
 \simeq\pi^2\nu D \frac {\ln^2[t\Delta(R/l)^2]} {\ln (R/l)}
\label{8}
\end{equation}

For the orthogonal and symplectic ensembles, the saddle-point equation
(\ref{1}) has the same form, with the only difference that the action
(\ref{5}) is multiplied by the factor $\beta/2$, where $\beta=1,2,4$
for the orthogonal, unitary and symplectic symmetries respectively.
Combining eqs.(\ref{5}) and (\ref{8}), we get thus for the long-time
asymptotics of the average
conductance $G(t)\sim e^{-S}$ in all three symmetry cases:
\begin{eqnarray}
&  G(t)\sim
(t\Delta)^{-2\pi\beta g}\ , & \quad 1\ll t\Delta\ll (R/l)^2 \label{9d}\\
& G(t)\sim \exp\left\{-{\pi\beta g\over 4} {\ln^2(t/g\tau)\over
\ln(R/l)}\right\}\ ,  &\quad
t\Delta\gg (R/l)^2
\label{9}
\end{eqnarray}
where $g=2\pi\nu D$ is the dimensionless conductance per square in 2D
and $\tau$ is the mean free time.

The far asymptotical behavior (eq.(\ref{9})) is of the
LN form and very similar to that found by AKL (see eq.(7.8) in
Ref.\cite{AKL}). It differs only by the factor $1/g$ in the argument
of $\ln^2$. It is easy to see however that this difference disappears if
one does the last step of the AKL calculation with a better accuracy.
Let us consider for this purpose the intermediate expression of AKL
(Ref.\cite{AKL}, eq.(7.11)):
\begin{equation}
G(t)\propto -{\sigma\over\tau}\int_0^\infty e^{-t/t_\phi}\exp\left[
-{1\over 4u}\ln^2{t_\phi\over\tau}\right]{dt_\phi\over t_\phi}
\label{10}
\end{equation}
where $u\simeq{1\over 2\pi^2\nu D}\ln{R\over l}$ in the weak
localization region in 2D, which we are considering. Evaluating the
integral (\ref{10}) by the saddle point method, we find
\begin{eqnarray}
G(t)&\sim& \exp\left\{-{1\over 4u}\ln^2{2ut\over\tau}\right\}
\nonumber\\
&\sim& \exp\left\{-{\pi g\over 4} {\ln^2(t/g\tau)\over
\ln(R/l)}\right\} \ ,
\label{11}
\end{eqnarray}
where we have kept only the leading term in the
exponent. Eq.(\ref{11}) is in {\it exact} agreement with
eq.(\ref{9}) for $\beta=1$ (AKL assumed the orthogonal
symmetry of the ensemble).
Therefore, the supersymmetric treatment confirms the AKL result and
also establishes the region of its validity. It is instructive to
represent the obtained results in terms of the superposition of simple
relaxation processes with mesoscopically distributed relaxation times
$t_\phi$:
\begin{equation}
G(t)\sim\int {dt_\phi\over t_\phi} e^{-t/t_\phi} P(t_\phi)
\label{11s}
\end{equation}
Then we have from eqs.(\ref{9d}), (\ref{9}) for the distribution
function $P(t_\phi)$:
\begin{equation}
P(t_\phi)\sim\left\{
\begin{array}{ll}
(t_\phi/t_D)^{-2\pi\beta g}\ , &\ \
t_D\ll t_\phi\ll t_D \left({R\over l}\right)^2 \\
\exp\left\{-{\pi\beta g\over 4} {\ln^2(t_\phi/\tau)\over \ln
(R/l)}\right\} \ ,&\ \ t_\phi\gg t_D \left({R\over l}\right)^2\ ,
\end{array}
\right.
\label{11t}
\end{equation}
where $t_D\simeq R^2/D$ is the time of diffusion through the sample.

For completness, we list also the results for quasi-1D and 3D systems.
For a quasi-1D sample (wire) of the length $L$ (which is assumed to be
much shorter than the localization length $\xi=2\beta\pi\nu D$)
the asymptotics read
\begin{equation}
G(t)\sim\exp\left\{-{\beta\pi\nu D\over L}\ln^2(t\Delta)\right\}\ ,\ \
t\Delta\gg 1
\label{12}
\end{equation}
(for $\beta=2$ this is just eq.(16) of MK). It is interesting to note
that eq.(\ref{12}) has essentially the same form as the asymptotical
formula for $G(t)$ found by Altshuler and Prigodin \cite{AP} for the
{\it strictly} 1D sample with a length much {\it exceeding}
the localization length:
\begin{equation}
G(t)\sim\exp\left\{-{l\over L}\ln^2(t/\tau)\right\}
\label{13}
\end{equation}
If we replace in eq.(\ref{13}) the 1D localization length $\xi=2l$ by the
quasi-1D localization length $\xi=2\beta\pi\nu D$, we  reproduce
the asymptotics (\ref{14}) (up to a normalization of $t$ in the
argument of
$\ln^2$, which does not affect the leading term in the exponent for
$t\to\infty$). This leads us to make the following two
conclusions. Firstly, this confirms once more the general conjecture
\cite{FM} that the statistical properties of smooth envelopes of the
wave functions in 1D and quasi-1D samples are identical. Secondly,
this shows that the asymptotical ``tail'' (\ref{12}) in the metallic
sample is indeed due to ``quasi-localized'' eigenstates, as has been
conjectured \cite{MK,AP,AP1,Kravtsov}.

In 3D, the analysis proceeds along the same line as for the
ultra-long-time region in 2D. This is essentially what has been done
by MK in their consideration of the 3D case. The result at
$t\gg(k_fl)^2t_D$ (where $k_f$ is the Fermi momentum) reads:
\begin{equation}
 G(t)\sim\exp\{-S(t)\}\ ,\qquad S(t)\sim (k_f l)^2\ln^3\left[{t\over
\tau(k_f l)^2}\right]
\label{14}
\end{equation}
In contrast to the 2D case, the exact numerical coefficient in the
exponent in eq.(\ref{14}) cannot be found within the diffusion
approximation.

We note in conclusion, that the obtained long-time asymptotics of the
average conductance have a very similar form to the asymptotical
behavior of the  distribution function $P(\rho)$
of local density of states (LDOS)
\cite{LDOS}. In both cases, the result is of the LN form in quasi-1D and
2D, and of a  somewhat different
(though very similar) $\exp\{-(k_f l)^2\ln^3(\cdot)\}$
 form in 3D. As in the case of LDOS distribution \cite{LDOS}, we have
found a perfect agreement with the result of renormalization group (RG)
treatment \cite{AKL} in 2D.  I believe this agreement between the RG
and supersymmetric treatments of $G(t)$ and $P(\rho)$ to be of
considerable conceptual importance. To make this point clear, I remind
the reader that one of the first achievments of the supersymmetry
method as applied to disordered electronic systems was the detailed
study of the Anderson metal--insulator transition on the
(effectively infinite-dimensional) Bethe
lattice \cite{BL}. The found non-power-law critical behavior seemed at
first sight to be in contradiction with the scaling hypothesis and
with the results of RG treatment. Since the solution of the Bethe
lattice problem was exact, this apparent contradiction questioned the
validity of the scaling and RG approaches. These doubts were supported
by the fact the solution in \cite{BL} heavily relied on the
non-compact structure of the supersymmetric $\sigma$-model manifold
and was dominated by the large values $\theta\gg 1$ of the
``non-compact angle'' $\theta$. On the other hand, the RG
consideration is just a resummation of the perturbative expansion and
does not distinguish between the compact and non-compact versions of
the $\sigma$-model.

However, we have been able to show recently \cite{BL-LDOS}
that the exotic critical behavior found in \cite{BL} is the property
of infinite-dimensional models only and transits to a power-law one
for a finite value $d<\infty$ of the space dimension, in qualitative
agreement with predictions of the scaling and RG approaches. Results
of \cite{LDOS} and of the present paper show a perfect quantitative
agreement of supersymmetry and RG methods when applied to the problem
of asymptotical behavior of various distributions in the ensemble of
mesoscopic metallic samples in the  weak localization region. This
provides strong support to other results obtained within the RG
approach in the weak localization region and in the vicinity of the
Anderson transition \cite{AKL}. On the other hand, we see that the
supersymmetry method is in many cases
able to reproduce  results of RG treatment
in a more elegant way. Furthermore, it is not restricted like RG
to the spatial dimension $d=2$ and can be
successfully applied to quasi-1D and 3D systems as well.
Besides the study of conductivity relaxation $G(t)$ and LDOS
distribution $P(\rho)$ discussed above, I would like to mention in
this context the recent progress in understanding of the statistical
properties of eigenfunctions \cite{FM,FM1,FE}. Seeing that the two
approaches are in amazingly good agreement, we can (depending on the problem
considered) use any of them or even combine them to complete our
understanding of the  properties of mesoscopic disordered systems.

I am grateful to V.E.Kravtsov and D.E.Khmelnitskii for valuable
discussions and comments.
This work was supported by SFB 195 der Deutschen Forschungsgemeinschaft.

\end{document}